\input harvmac
%
\def\journal#1&#2(#3){\unskip, \sl #1\ \bf #2 \rm(19#3) }
\def\andjournal#1&#2(#3){\sl #1~\bf #2 \rm (19#3) }

\def\ie{{\it i.e.}}
\def\eg{{\it e.g.}}

\def\frac#1#2{{#1\over#2}}

\def\d{\partial}

\def\inbar{\,\vrule height1.5ex width.4pt depth0pt}
\def\IC{\relax\hbox{$\inbar\kern-.3em{\rm C}$}}
\def\IR{\relax{\rm I\kern-.18em R}}
\def\IP{\relax{\rm I\kern-.18em P}}

%
%

%
\catcode`\@=11
\def\slash#1{\mathord{\mathpalette\c@ncel{#1}}}
\overfullrule=0pt

\def\MM{{\cal M}}
\def\NN{{\cal N}}

\def\underrel#1\over#2{\mathrel{\mathop{\kern\z@#1}\limits_{#2}}}

\catcode`\@=12


%


\def\ads{{AdS_3}}
\def\slr{{SL(2)}}
\def\ul{{U(1)}}
\def\nul{{\NN/\ul}}

\def\js{{\bf J}}
\def\gs{{\bf G}}
\def\ls{{\bf L}}

\def\e{\epsilon}

\def\[{[}
\def\]{]}

\def\comment#1{ }

\nref\gks{A.\ Giveon, D.\ Kutasov, and N.\ Seiberg,
``Comments on String Theory on $\ads$," Adv. Theor. Math.
Phys. {\bf 2} (1998) 733, hep-th/9806194.}
\nref\bort{J.\ de Boer, H.\ Ooguri, H.\ Robins, and J.\ Tannenhauser,
``String Theory on $\ads$," JHEP 9812 (1998) 026, hep-th/9812046.}
\nref\ks{D.\ Kutasov and N.\ Seiberg, ``More Comments on String Theory on
$\ads$," hep-th/9903219.}
\nref\efgt{S.\ Elitzur, O.\ Feinerman, A.\ Giveon, and D.\ Tsabar, ``String
Theory on $\ads\times S^3\times S^3\times S^1$," hep-th/9811245.}
\nref\kll{D.\ Kutasov, F.\ Larsen, and R.\ G.\ Leigh, ``String
Theory in Magnetic Monopole Backgrounds," hep-th/9812027.}
\nref\more{S.\ Yamaguchi, Y.\ Ishimoto, and K.\ Sugiyama, ``$\ads/CFT_2$
Correspondence and Space-Time $N=3$ Superconformal Algebra," hep-th/9902079.}
\nref\sw{N.\ Seiberg and E.\ Witten, ``The D1/D5 System and Singular CFT,"
hep-th/9903224.}
\nref\oldrefs{
A. B. Zamolodchikov and V. A. Fateev, Sov. J. Nucl. Phys. {\bf 43}
(1986) 657;
J. Balog, L. O'Raifeartaigh, P. Forgacs, and A. Wipf,
Nucl. Phys. {\bf B325} (1989) 225;
L. J. Dixon, M. E. Peskin and J. Lykken,  Nucl. Phys. {\bf B325} (1989)
329;
A. Alekseev and S. Shatashvili, Nucl. Phys. {\bf B323} (1989) 719;
N. Mohameddi, Int. J. Mod. Phys. {\bf A5} (1990) 3201;
P. M. S. Petropoulos, Phys. Lett. {\bf B236} (1990) 151;
M. Henningson and S. Hwang, Phys. Lett. {\bf B258} (1991) 341;
M. Henningson, S. Hwang, P. Roberts, and B. Sundborg, Phys. Lett. {\bf
B267} (1991) 350;
S. Hwang, Phys. Lett. {\bf B276} (1992) 451, hep-th/9110039;
I. Bars and D. Nemeschansky, Nucl. Phys. {\bf B348} (1991) 89;
S. Hwang, Nucl. Phys. {\bf B354} (1991) 100;
K. Gawedzki, hep-th/9110076;
I. Bars, Phys. Rev. {\bf D53} (1996) 3308, hep-th/9503205;
in {\it Future Perspectives In String Theory} (Los Angeles, 1995),
hep-th/9511187;
O. Andreev, hep-th/9601026, Phys.Lett. {\bf B375} (1996) 60.
J. L. Petersen, J. Rasmussen and M. Yu, hep-th/9607129, Nucl.Phys. {\bf
B481} (1996) 577;
Y. Satoh, Nucl. Phys. {\bf B513} (1998) 213, hep-th/9705208;
J. Teschner, hep-th/9712256,  hep-th/9712258;
J. M. Evans, M. R. Gaberdiel, and M. J. Perry, hep-th/9806024;
hep-th/9812252.}
\nref\bdfm{T.\ Banks, L.\ Dixon, D.\ Friedan, and E.\ Martinec,
``Phenomenology and Conformal Field Theory or Can String Theory Predict
the Weak Mixing Angle?" Nucl.\ Phys.\ {\bf B299} (1988) 613.}
\nref\bd{T.\ Banks and L.\ Dixon, ``Constraints on String Vacua with
Spacetime Supersymmetry," Nucl.\ Phys.\ {\bf B307} (1988) 93.}
\nref\sv{S.\ Shatashvili and C.\ Vafa, ``Superstrings and Manifolds of
Exceptional Holonomy," Selecta Math. {\bf A1} (1995) 347, hep-th/9407025.}
\nref\jmf{J.\ Figueroa-O'Farrill, ``A note on the extended
superconformal algebras associated with manifolds of
exceptional holonomy," Phys.\ Lett.\ {\bf B392} (1997) 77, hep-th/9609113.}

\rightline{RI-3-99}
\rightline{ITP-SB-99-10}
\Title{
\rightline{hep-th/9904024}}
{\vbox{\centerline{Supersymmetric string vacua on $AdS_3 \times \NN$ }}}
\medskip
\centerline{Amit Giveon\footnote{$^*$}{e-mail address:
giveon@vms.huji.ac.il}}
\smallskip
\centerline{\it Racah Institute of Physics, \it The Hebrew University}
\centerline{\it Jerusalem 91904, Israel}
\bigskip
\centerline {and}
\medskip
\centerline {Martin Ro\v cek\footnote{$^{**}$}{e-mail address:
rocek@insti.physics.sunysb.edu}}
\centerline{\it Institute of Theoretical Physics,
State University of New York }
\centerline{\it Stony Brook, NY 11794-3840, USA }
\bigskip\bigskip\bigskip
\noindent
String backgrounds of the form $\ads\times\NN$ that give
rise to two dimensional spacetime superconformal symmetry are constructed.

\vfill

\Date{4/99}
\newsec{Introduction}

In this letter we study the conditions on curved string backgrounds of the
form $\ads\times\NN$ that give rise to spacetime superconformal symmetry. We
use the NSR formulation; for simplicity we describe the left moving sector
only -- it can be combined with the right-moving sector in the standard way.
String theories on $\ads\times\NN$ were studied for the bosonic case in
\refs{\gks,\bort,\ks}. Examples of supersymmetric strings in the NSR
formulation were studied in, \eg, \refs{\gks,\efgt,\kll,\more,\ks,\sw}. 
For some early work on string theory on $\ads$ see, \eg, \oldrefs.

The main results of this work are the following: If $\NN$ has an affine
$\ul$ symmetry and $\nul$ has an $N=2$ worldsheet superconformal symmetry,
then there is a construction of a superstring with two-dimensional $N=2$
$spacetime$ superconformal symmetry. A $Z_2$ quotient of this construction
leads to a family of theories with two-dimensional $N=1$ $spacetime$
superconformal symmetry. We also discuss conditions for $N>2$ superconformal
symmetry: These involve an $\NN$ with an $SU(2)$ factor whose level
is determined in terms of the level of the $\ads$ background.

This investigation is the analog of the study of supersymmetric backgrounds
for compactification to Minkowski space $\MM_d$ in $d=3$ or $4$ dimensions
\refs{\bdfm,\bd,\sv,\jmf}. String theories on $\MM_4\times\NN$ have
four-dimensional $N=1$ spacetime supersymmetry provided $\NN$ has an $N=2$
worldsheet superconformal symmetry \refs{\bdfm,\bd}.  String theories on
$\MM_3\times\NN$ have three-dimensional $N=2$ spacetime supersymmetry if
$\NN$ has an affine $\ul$ and $\nul$ has an $N=2$ worldsheet superconformal
symmetry. A $Z_2$ quotient of such an $\NN$ leaves  a three-dimensional $N=1$
spacetime supersymmetry; in this case, the symmetry algebra on $\NN$ can be
extended to a nonlinear algebra associated to manifolds with $G_2$ holonomy
\refs{\sv,\jmf}.

The structure of the paper is as follows: In section 2, we describe the
worldsheet properties that lead to spacetime supersymmetry on
$\ads\times\NN$. In section 3, we construct the two dimensional $N=2$
spacetime superconformal algebra associated with the boundary CFT of $\ads$.
In section 4, we take a quotient of the $N=2$ construction to find a class
of models with $N=1$ spacetime superconformal symmetry. In section 5, we
discuss models with $N>2$ spacetime superconformal symmetry. Finally, in
section 6, we comment on new examples that arise from our results and other
issues.

\newsec{Worldsheet properties of fermionic strings on $AdS_3\times \NN$}

We first consider the $\ads$ factor of the background.
This theory has affine $\slr$ currents
\eqn\scur{\psi^A+\theta \sqrt{\frac2k} J^A ~,\qquad A=1,2,3,}
where
\eqn\jA{J^A=j^A-\frac{i}{2}\epsilon^A{}_{BC}\psi^B \psi^C~,}
and
\eqn\norms{\eqalign{\psi^A(z)\psi^B(w)
&\sim\frac{\eta^{AB}}{z-w}~,
\qquad \eta^{AB}={\rm diag}(+,+,-)~, \cr &\cr
J^A(z)J^B(w)
&\sim\frac{\frac{k}2\eta^{AB}}{(z-w)^2}+
\frac{i\epsilon^{AB}{}_CJ^C}{z-w}~.}}
The purely bosonic currents $j^A$ generate an affine $\slr$ algebra at level
$k+2$ and commute with $\psi^A$, whereas the total currents $J^A$ generate a
level $k$ $\slr$ algebra and act on $\psi$ as follows from \jA,\norms.
The central charge of the $\ads$ part of the theory is thus
\eqn\cslr{c^\slr=\frac{3(k+2)}k+\frac32~,}
where the two terms are the bosonic and fermionic contributions,
respectively. The $N=1$ worldsheet supercurrent is
\eqn\tfslr{T_F^\slr=\sqrt{\frac2k}(\psi^Aj_A-i\psi^1\psi^2\psi^3)~.}

The internal space $\NN$ is described by a unitary superconformal field
theory (CFT) background with central charge
\eqn\cnn{c^\NN=15-c^\slr=\frac{21}2-\frac6k~.}
We denote the worldsheet supercurrent of $\NN$ by $T_F^\NN$.

The construction
of $N=2$ spacetime supersymmetry described in section 3 below requires that
$\NN$ have an affine $\ul$ symmetry with an $N=1$ current
\eqn\ucur{\psi^\ul+\theta J^\ul~,}
where
\eqn\unorm{\eqalign{\psi^\ul(z)\psi^\ul(w)
&\sim\frac1{z-w}~, \cr &\cr
J^\ul(z)J^\ul(w)
&\sim\frac1{(z-w)^2}~, \cr &\cr
J^\ul(z)\psi^\ul(w) &\sim 0~,}}
and a worldsheet supercurrent
\eqn\scul{T_F^\ul=\psi^\ul J^\ul~.}
It is convenient to bosonize the affine current
\eqn\bul{J^\ul=i\d Y}
where $Y$ is a canonically normalized scalar: $Y(z)Y(w)\sim-\log(z-w)$.

We can construct the quotient CFT $\NN/\ul$ with the supercurrent
\eqn\qtf{T^{\nul}_F=T^\NN_F-T^\ul_F~;}
this has a central charge
\eqn\cqft{c^{\nul}=c^\NN-c^\ul=9-\frac6k~.}
The construction of $N=2$ spacetime supersymmetry described in section 3
below further requires that $\nul$ have an $N=2$ superconformal algebra
(which commutes with the $\ul$ above). In particular, its $\ul_R$-current
$J^\nul_R$ has the standard normalization
\eqn\rnorm{J^\nul_R(z)J^\nul_R(w) \sim\frac13\,\frac{c^\nul}{(z-w)^2}~.}
We bosonize $J^\nul_R$ in terms of a canonically normalized scalar $Z$ by
\eqn\br{J^\nul_R=i\sqrt{\frac{c^\nul}3}\d Z\equiv ia\d Z~,}
where
\eqn\asq{a\equiv\sqrt{3-\frac2k}~.} The worldsheet supercurrent $T^\nul_F$
can be decomposed into two parts with $R$-charges $\pm1$; these charges can
be expressed in terms of explicit $Z$ dependent factors to give:
\eqn\tfnulpm{T^\nul_F=e^{\frac{i}aZ}\tau_++ e^{-\frac{i}aZ}\tau_-~,}
where $\tau_\pm$ carry $no$ $R$-charge, \ie
\eqn\jract{\eqalign{J^\nul_R(z)e^{\pm\frac{i}aZ}(w) &\sim\frac{\pm
e^{\pm\frac{i}aZ}}{z-w}\cr &\cr
J^\nul_R(z)\tau_\pm(w)&\sim 0~.}}

\newsec{$N=2$ spacetime superconformal theories}

We now construct an $N=2$ superconformal algebra in
spacetime out of the worldsheet ingredients described above.
As in \gks, we introduce canonically normalized scalars $H_I$ with $I=0,1,2$:
\eqn\hint{\eqalign{\d H_1 & =\psi^1\psi^2\cr i\d H_2 & =\psi^3\psi^\ul\cr
i\sqrt3\d H_0 & =J^\nul_R-\sqrt{\frac2k}J^\ul~,}}
where
\eqn\hnorm{H_I(z)H_J(w)\sim-\delta_{IJ}\log(z-w)~.}
For future reference we remind the reader that
\eqn\basic{e^{\pm iH_2}=\frac{i}{\sqrt2}(\psi^3\pm\psi^\ul)~.}
The {\it spacetime} supercharges are constructed as
\ref\fms{D.\ Friedan, E.\ Martinec, and S.\ Shenker, ``Conformal
Invariance, Supersymmetry, and String Theory," Nucl.\ Phys.\ {\bf B271}
(1986) 93.}
\eqn\gspace{\gs^\pm_r=(2k)^\frac14\oint dz\,
e^{-\frac\phi2} S^\pm_r~,\qquad r=\pm\frac12~,}
where $\phi$ is the scalar field arising in the bosonized $\beta,\gamma$
superghost system of the worldsheet supersymmetry, and the spin
fields $ S^\pm_r$ are (recall \hint,\asq,\br,\bul)
\eqn\splus{\eqalign{ S^+_r & =e^{ir(H_1+H_2)+i\frac{\sqrt3}2H_0}=
e^{ir(H_1+H_2)+i\frac{a}2Z-i\sqrt{\frac1{2k}}Y}~,\cr
 S^-_r & =e^{ir(H_1-H_2)-i\frac{\sqrt3}2H_0}=
e^{ir(H_1-H_2)-i\frac{a}2Z+i\sqrt{\frac1{2k}}Y}~.}}
(We neglect the usual cocycle factors).
The supercharges $\gs^\pm_r$ are physical only if they are
BRST invariant. This requires that the OPE of $T_F(z) S^\pm_r(w)$ have no
$(z-w)^{-3/2}$ singularity. Here $T_F$ is the total worldsheet $N=1$
supercurrent:
\eqn\ttf{T_F=T^\slr_F+T^\ul_F+T^\nul_F}
(see \tfslr,\scul,\qtf).
Consider
\eqn\seee{ S_{\e_1\e_2\e}=
e^{\frac{i}2(\e_1H_1+\e_2H_2+\e(aZ-\sqrt{\frac2k}Y))}~,
\qquad\e_1,\e_2,\e=\pm1~;}
{}From \tfnulpm\ we find
\eqn\tsope{T^\nul_F(z)S_{\e_1\e_2\e}(w)\sim
(z-w)^{\frac\e2}(...)\tau_++(z-w)^{-\frac\e2}(...)\tau_-~,}
where $(...)$ represents irrelevant factors.
Therefore $T^\nul_F(z) S^\pm_r(w)$ has $no$ $(z-w)^{-3/2}$
singularity, and the only possible sources of such singularities are
$T^\ul_F$ and the $\psi^1\psi^2\psi^3$ term in $T^\slr_F$. These two
contributions cancel each other for $\e_1\e_2\e=1$, as can be seen using
\eqn\trel{\psi^\ul J^\ul -i \sqrt{\frac2k}\psi^1\psi^2\psi^3=
(\frac1{\sqrt2}\d Y-\frac1{\sqrt k}\d H_1)e^{iH_2}-(\frac1{\sqrt2}\d
Y+\frac1{\sqrt k}\d H_1)e^{-iH_2}}
(see \bul,\hint,\basic). Substituting all 4 solutions of
$\e_1\e_2\e=1$  into \seee, we recover \splus. In addition, 
$e^{-\phi/2} S^\pm_r$ are mutually local.
This completes the proof that
$\gs^\pm_r$ as defined in \gspace\ are physical.

The algebra generated by the supercharges is
\eqn\stalg{\eqalign{\{\gs^+_r,\gs^-_s\} & =2\ls_{r+s}+(r-s)\js_0
~,\qquad r,s=\pm\frac12\cr
\[\ls_m,\ls_n\] & = (m-n)\ls_{m+n}~,~~~\,\qquad m,n=0,\pm1 \cr
\[\ls_m,\gs^\pm_r\] & = (\frac{m}2-r)\gs^\pm_{m+r} \cr
\[\js_0,\gs^\pm_r\] & =\pm \gs^\pm_r}}
with all other (anti)commutators vanishing. Up to picture-changing \fms,
$\ls_0,\ls_{\pm1},\js_0$ are given by (recall \scur,\ucur)
\eqn\stvir{\ls_0=-\oint J^3~,\qquad
\ls_{\pm1}=-\oint \frac1{\sqrt2}(J^1\pm iJ^2)~,}
\eqn\stul{\js_0=-\sqrt{2k}\oint J^\ul~.}
The algebra \stalg\ is a global spacetime $N=2$ superconformal algebra.

String theory on $\ads\times\NN$ has a $full$ spacetime Virasoro symmetry
$\ls_n$, with $n\in{\bf Z}$ \refs{\gks,\ks}; commuting $\ls_n$ with the
generators of the global algebra \stalg\ gives a full spacetime $N=2$
superconformal algebra in the spacetime NS-sector with modes $\gs^\pm_r$,
$r\in{\bf Z}+\frac12$ and $\js_n$, $n\in{\bf Z}$. Physical states are
constructed using physical vertex operators that are local with respect to
the supercharges \gspace; this is the analog of the usual $GSO$
projection\foot{Strictly speaking, the full Virasoro
algebra and physical states were constructed in the Euclidean version
of $\ads$ \refs{\gks,\ks}. The construction of the finite Lie superalgebra
\stalg\ uses only the algebraic structure of $SL(2)$, and is independent of
the representation theory; hence it is also valid for the Lorentzian case.}.

\newsec{$N=1$ spacetime superconformal theories}
The construction of the previous section gave us two dimensional $N=2$
spacetime superconformal symmetry. It is straightforward to find a $Z_2$
quotient that preserves exactly half of the spin fields \splus\ and leads to
$N=1$ spacetime superconformal symmetry.  This quotient is analogous to the
construction of manifolds with $G_2$ holonomy by a $Z_2$ quotient of a product
of a Calabi-Yau manifold with an $S^1$ \jmf.

Concretely, we break the $N=2$ superconformal symmetry of $\nul$ by
the quotient with respect to $J_R^\nul\to-J_R^\nul$; simultaneously, we
take the quotient with respect to $J^\ul\to-J^\ul$ and $\psi^\ul\to-\psi^\ul$.
This has the net effect of identifying $\{H_1,H_2,H_0\}\to\{H_1,-H_2,-H_0\}$
(see \hint), and thus $S^\pm_r\to S^\mp_r$ \splus.  Therefore, the spacetime
superconformal symmetry is projected to the $N=1$ subalgebra generated by the
symmetric combination $\gs^+_r+\gs^-_r$.

This indeed resembles the construction of superconformal models on manifolds
with $G_2$ holonomy \refs{\sv,\jmf}, except that the total central charge of
$\NN$ is not $21/2$ but rather $21/2-6/k$ (see \cnn). It would be interesting
to see if one can find a general nonlinear worldsheet algebra that
characterizes $\NN$ and then use the methods of \sv\ to generate the $N=1$
spacetime superconformal symmetry in the general $\ads\times\NN$ case.

\newsec{$N>2$ spacetime superconformal theories}

We may also consider the extension of the methods of section 3 to models with
$N>2$ symmetry.  This gives rise to models that have been considered on a case
by case basis in the literature \refs{\gks,\efgt,\kll,\more}.

The small $N=4$ superconformal algebra (see \gks\ and references therein) has
an affine $SU(2)$ $R$-symmetry.  As explained in \gks, this spacetime affine
$SU(2)$ arises from a level $k$ worldsheet affine $SU(2)$ factor in $\NN$.
For the construction of section 3, we may take $J^\ul$ as the Cartan
generator of $SU(2)_k$.  The remaining background
$\NN_{(c=6)}\equiv\NN/SU(2)_k$ is precisely a $c=6$ CFT, and small
$N=4$ spacetime supersymmetry requires that $\NN_{(c=6)}$ have small $N=4$
worldsheet supersymmetry.  This can be shown by the methods  in \bd, where it
is argued that for compactification to four dimensional Minkowski space
$\MM_4\times\NN_{(c=9)}$, $N=2$ spacetime supersymmetry on $\MM_4$ requires
that $\NN_{(c=9)}$ factorize as $\NN_{(c=9)}=\NN_{(c=6)}\times T^2$ with
small $N=4$ worldsheet superconformal symmetry on $\NN_{(c=6)}$.

The large $N=4$ superconformal algebra (see \efgt\ and references therein) has
an affine $SU(2)\times SU(2)\times\ul$ $R$-symmetry. Again, as explained in
\refs{\gks,\efgt}, this spacetime affine algebra arises from a worldsheet
algebra $SU(2)_{k'}\times SU(2)_{k''}\times\ul$ where the levels are related
to the level $k$ of the $\ads$ factor by $1/k=1/k'+1/k''$. This implies that
the central charge of the $SU(2)_{k'}\times SU(2)_{k''}\times\ul$ factor is
$c=21/2-6/k$, and so completely determines $\NN$ \cnn. For the construction
of section 3, we may take $J^\ul$ as the diagonal Cartan generator of
$SU(2)_{k'}\times SU(2)_{k''}$; this implies that $H_2$ of \hint\ above is
$-H_4$ of equation (2.31) in \efgt.

To construct $N=3$ spacetime superconformal models, we may for instance take
a $Z_2$ quotient of the large $N=4$ model in such a way as to preserve 3 out
of 4 spacetime supersymmetries.  This is worked out in detail in
\more; the basic idea is to take the construction of \efgt\ with $k'=k''$ and
quotient by a $Z_2$ action that exchanges the two $SU(2)$ factors in
$\NN$ and simultaneously reflects the $\ul$ factor in $\NN$. Since the
$J^\ul$ current we use is in the diagonal of $SU(2)\times SU(2)$  and hence
inert under this quotient, the construction survives and gives an $N=2$
subalgebra of the $N=3$ spacetime superconformal algebra discussed in \more.

In models with enhanced spacetime superconformal symmetries, one has to take
some care in choosing $J^\ul$, as an arbitrary choice may lead to
supercharges that are not mutually local with the spacetime $R$-symmetries,
and thus preserve only the $N=2$ subalgebra \stalg.

\newsec{Examples and discussion}

We close with a few remarks:

\item{1.}
The construction of section 3 can be used to find many new examples of
$\ads\times\NN$ string backgrounds with spacetime superconformal
symmetry.  A broad class is given by $\NN=\ul\times\NN_{KS}$ where
$\NN_{KS}$ is a Kazama-Suzuki model \ref\ks{Y.\ Kazama and H.\ Suzuki,
``New $N=2$ Superconformal Field Theories and Superstring Compactification,"
Nucl.\ Phys.\ {\bf B321} (1989) 232.} with central charge $c=9-6/k$.
Kazama-Suzuki models are gauged $N=1$ WZW models $G/H$ with an enhanced
$N=2$ worldsheet superconformal symmetry. The cases
$(SU(2)_k\times\ul^4)/\ul$ or $(SU(2)_{k'}\times SU(2)_{k''}\times\ul)/\ul$
are precisely the cases with enhanced spacetime superconformal symmetry
discussed above. A simple new case is, for instance, $SU(3)_{4k}/\ul^2$.
\item{2.}
When the background has an enhanced worldsheet affine algebra, the
construction of section 3 can be generalized; in particular, if the enhanced
algebra includes an extra affine $\ul^2$ factor, $N>2$ spacetime symmetries 
can be constructed as in \efgt.
\item{3.}
The construction we have given here leads to $conformal$ spacetime
supersymmetries of the boundary CFT of $\ads$. Other constructions of 
spacetime supersymmetry are possible, such as the construction with respect
to the $\ul_R$ of the total worldsheet $N=2$ superconformal symmetry of
$\ads\times\NN$ (see, \eg, appendix B of \gks). These in general correspond
to different string vacua defined on the same $\sigma$-model background,
and do $not$ give rise to spacetime conformal symmetries\foot{Technically,
the theories differ because the physical states are required to be
mutually local with respect to different spin fields. The spin fields
\gspace\ are mutually local with respect to the $\slr$ currents, whereas the
spin fields in the appendix B of \gks\ are not.}. It would be interesting to
know if the construction given here is the unique one that does lead to
spacetime conformal symmetry (modulo the ambiguity noted in the previous
paragraph for spaces with $\ul^2$ factors).

\bigskip
\noindent{\bf Acknowledgements:}
We are happy to thank D.\ Kutasov for comments on the manuscript.
This work is supported in part by the BSF -- American-Israel Bi-National
Science Foundation. The work of AG is supported in part by the Israel
Academy of Sciences and Humanities -- Centers of Excellence Program. The work
of MR is supported in part by  NSF grant No.\ PHY9722101. AG thanks the ITP
at Stony Brook and MR thanks the Racah Institute at the Hebrew University for
their hospitality.

\listrefs
\end